\documentclass[conference]{IEEEtran}
\usepackage{cite}
\usepackage{amsmath,amssymb,amsfonts}
\usepackage{algorithmic}
\usepackage{graphicx}
\usepackage{textcomp}
\usepackage{xcolor}
\def\BibTeX{{\rm B\kern-.05em{\sc i\kern-.025em b}\kern-.08em
    T\kern-.1667em\lower.7ex\hbox{E}\kern-.125emX}}
\usepackage{mathtools}
\usepackage{dsfont}
\usepackage{tikz}
\usepackage{pgfplots}
\usepackage{subcaption}
\usepackage{booktabs}
\pgfplotsset{compat=newest}
\pgfplotsset{plot coordinates/math parser=false}
\newlength\figureheight
\newlength\figurewidth
\usepackage[nolist]{acronym}

\usepackage{textcomp}
\makeatletter
 \let\old@ps@headings\ps@headings
 \let\old@ps@IEEEtitlepagestyle\ps@IEEEtitlepagestyle
 \def\confheader#1{%
 \def\ps@headings{%
 \old@ps@headings%
 \def\@oddhead{\strut\hfill#1\hfill\strut}%
 \def\@evenhead{\strut\hfill#1\hfill\strut}%
 }%
 \def\ps@IEEEtitlepagestyle{%
 \old@ps@IEEEtitlepagestyle%
 \def\@oddhead{\strut\hfill#1\hfill\strut}%
 \def\@evenhead{\strut\hfill#1\hfill\strut}%
 }%
 \ps@headings%
 }
 \makeatother

\confheader{%
PUBLISHED IN MDPI ENERGIES, 28TH DECEMBER 2024, DOI: 10.3390/en18010084
 }

\begin{document}

\title{Thermal Finite-Element Model of an Electric Machine Cooled by a Spray\\
\thanks{This work is funded by the Deutsche Forschungsgemeinschaft (DFG, German Research Foundation) – Project-ID 492661287 – TRR 361, the Athene Young Investigator Program of TU Darmstadt, the Graduate School Computational Engineering at TU Darmstadt.}
}

\author{\IEEEauthorblockN{Christian Bergfried}
\IEEEauthorblockA{\textit{Institute for Accelerator Science}\\
\textit{ and Electromagnetic Fields (TEMF)} \\
\textit{Graduate School CE}\\
\textit{Technische Universität Darmstadt}\\
Darmstadt, Germany \\
christian.bergfried@tu-darmstadt.de}
\and
\IEEEauthorblockN{Samaneh Abdi Qezeljeh}
\IEEEauthorblockA{\textit{Institute for Fluid Mechanics}\\
\textit{and Aerodynamics} \\
\textit{Graduate School CE}\\
\textit{Technische Universität Darmstadt}\\
Darmstadt, Germany\\
abdi.qezeljeh@sla.tu-darmstadt.de}
\and
\IEEEauthorblockN{Ilia V. Roisman}
\IEEEauthorblockA{\textit{Institute for Fluid Mechanics}\\
\textit{and Aerodynamics}\\
\textit{Technische Universität Darmstadt}\\
Darmstadt, Germany}
\and
\IEEEauthorblockN{Herbert De Gersem}
\IEEEauthorblockA{\textit{Institute for Accelerator Science}\\
\textit{ and Electromagnetic Fields (TEMF)} \\
\textit{Graduate School CE}\\
\textit{Technische Universität Darmstadt}\\
Darmstadt, Germany}
\and
\IEEEauthorblockN{Jeanette Hussong}
\IEEEauthorblockA{\textit{Institute for Fluid Mechanics}\\
\textit{and Aerodynamics} \\
\textit{Graduate School CE}\\
\textit{Technische Universität Darmstadt}\\
Darmstadt, Germany}
\and
\IEEEauthorblockN{Yvonne Späck-Leigsnering}
\IEEEauthorblockA{\textit{Institute for Accelerator Science}\\
\textit{ and Electromagnetic Fields (TEMF)} \\
\textit{Graduate School CE}\\
\textit{Technische Universität Darmstadt}\\
Darmstadt, Germany}
}

\maketitle

\begin{abstract}
The demand for higher power density in electrical machines necessitates advanced cooling strategies. Spray cooling emerges as a promising and relatively straightforward technology, albeit involving complex physics. In this paper, a quasi-3D thermal finite-element model of a stator winding is created, by extrusion of a 2D cross-sectional finite-element model along the winding direction. The cooling effects of spray impact are simulated as a heat flux using an impedance boundary condition at the surface of the winding overhang. The results confirm the advantageous performance of spray cooling, indicating that it may enable a tenfold increase in power density compared to standard air- or water-cooled machines.
\end{abstract}


\section{Introduction}
\begin{figure}
    \centering
    \includegraphics[width=0.75\linewidth]{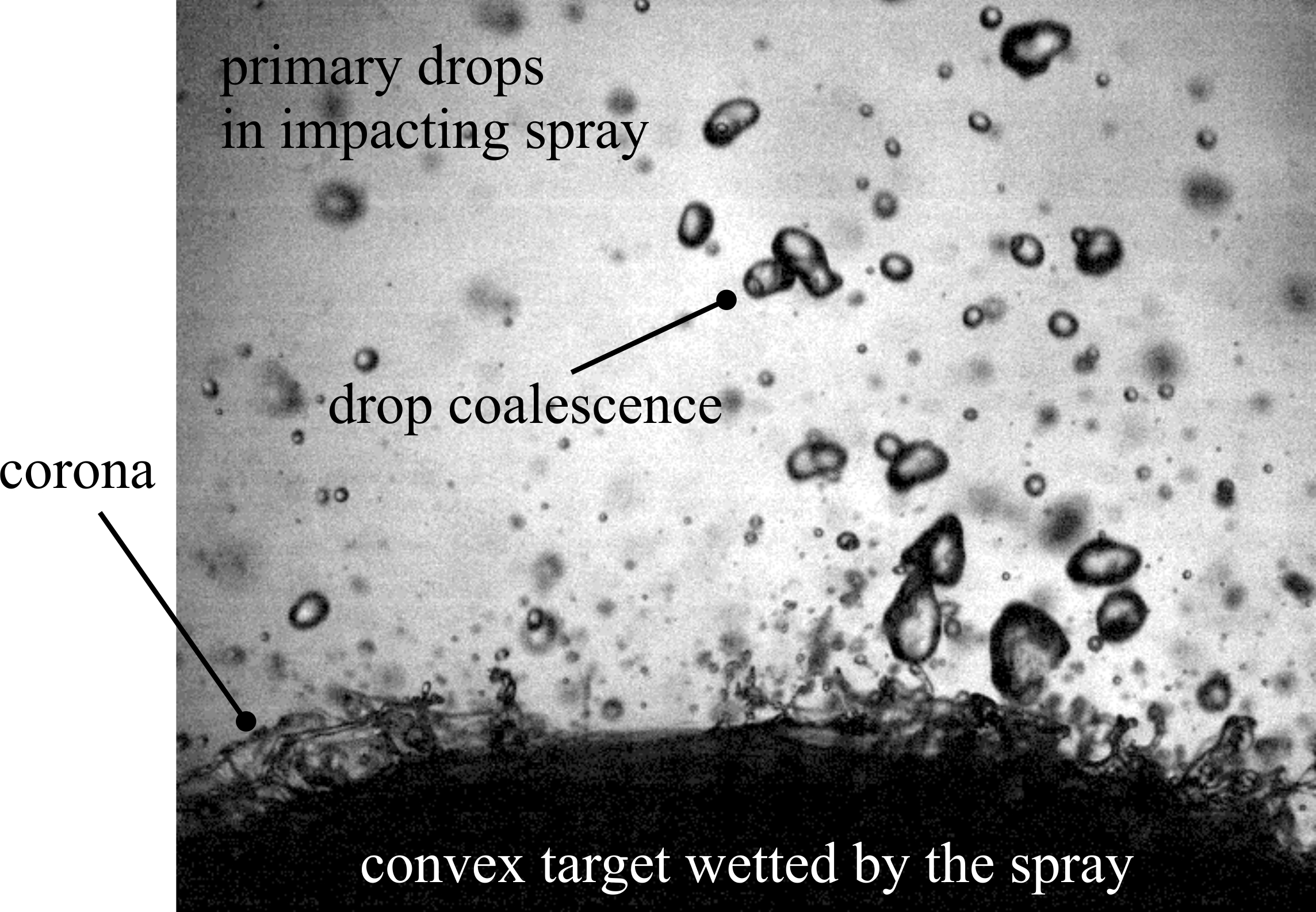}
    \caption{Shadowgraphy image of spray impact onto a convex solid  target.}
    \label{fig:spray_impact}
\end{figure}

\begin{figure}
    \centering
    \includegraphics[width=0.9\linewidth]{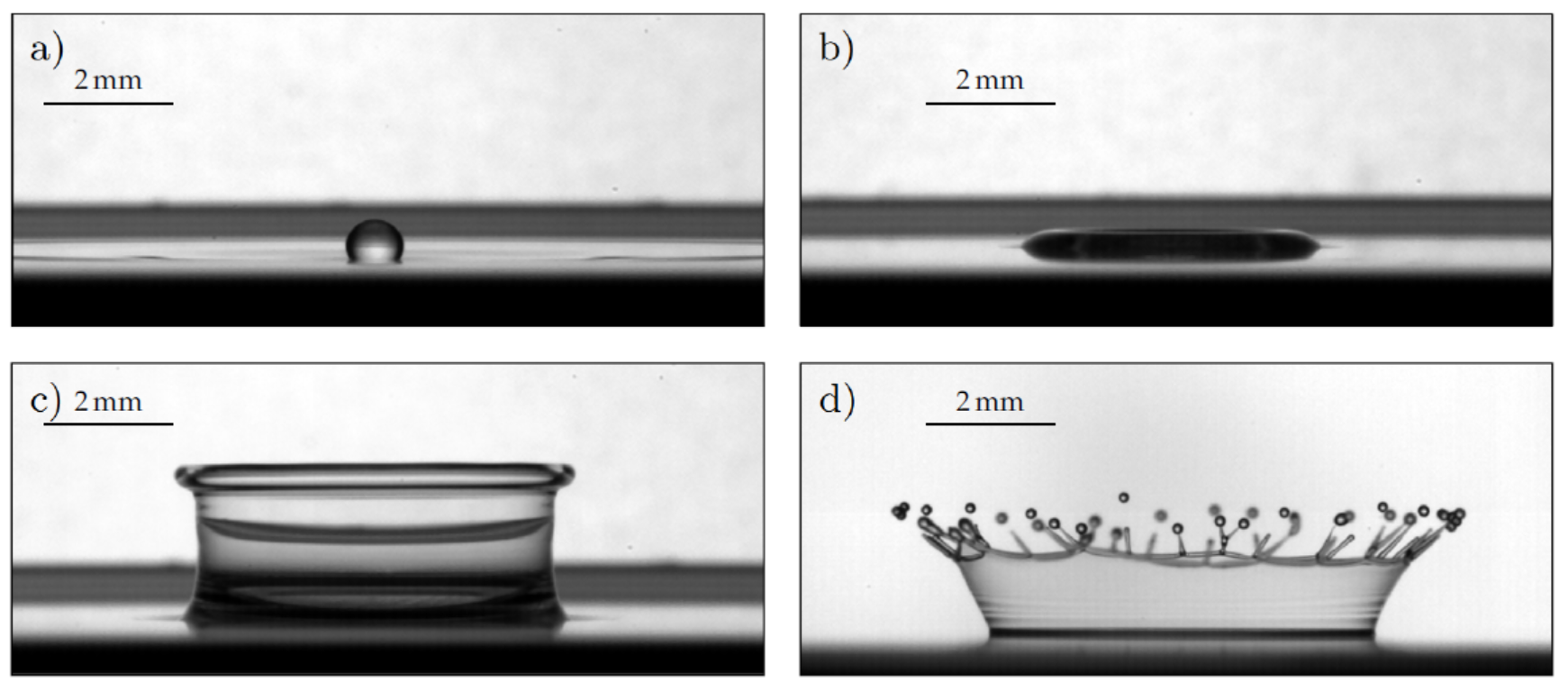}
    \caption{Typical outcomes of drop impact onto a wetted substrate. a) floating/bouncing (We = 0.62, Re = 24, $\delta =$ 0.33), b) deposition (We = 78, Re = 300,  $\delta =$ 0.33) c) crown formation (We = 700, Re = 900,  $\delta =$ 0.33, d) splash (We = 1224, Re = 604,  $\delta =$ 0.04).}
    \label{fig:dropoutcomes}
\end{figure}

\begin{figure}
    \centering
    \includegraphics[width=1\linewidth]{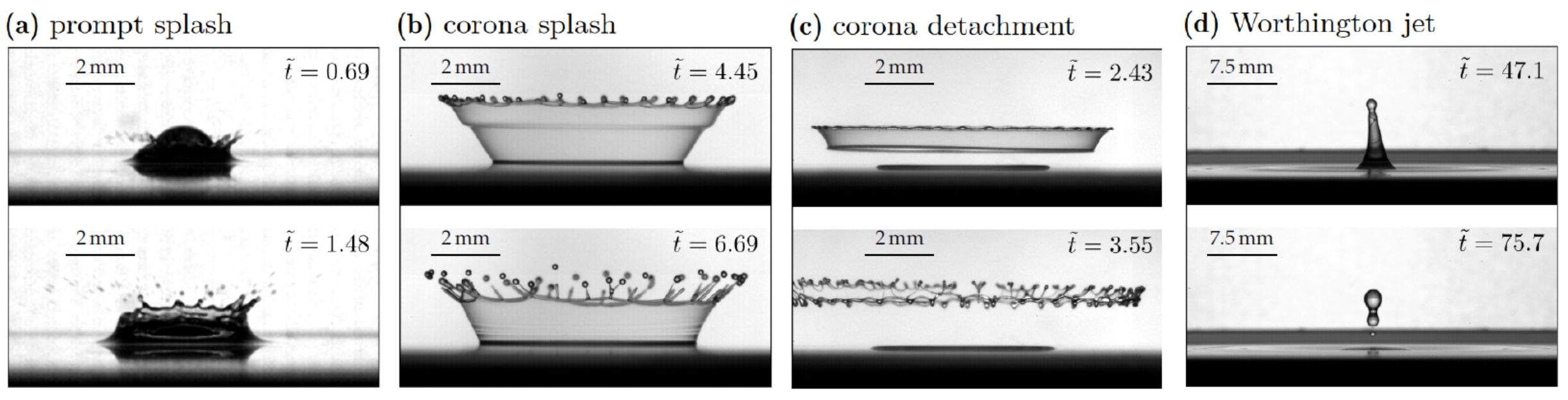}
    \caption{Various types of splash observed in experiments: a) prompt splash, b) corona splash, c) corona detachment, d) Worthington jet.}
    \label{fig:splashes}
\end{figure}

Electric machines produce heat due to electrical losses and mechanical friction. These losses are especially pronounced during startup or dynamic braking and tend to  increase significantly with heavier loads. Excessive heat can reduce machine efficiency and eventually lead to failure. This is why proper cooling of the electric machine parts is necessary to ensure that the machine operates at optimal temperatures, preventing overheating, which can lead to possible subsequent damage. Overheating can harm electronic components and shorten the machine’s lifespan.  High-speed electric machines face challenges due to higher loss densities. A thorough thermal analysis, combined with electromagnetic and mechanical considerations, as well as appropriate cooling processes, is essential during design \cite{rehman2018three,fan2010thermal,chong2021review}. Several traditional cooling methods, such as water-cooling jackets, oil, and air cooling, have been widely investigated in automotive applications \cite{7194542,en16197006}. 
Efficient cooling of high-power density electric machines is a rather challenging problem \cite{9819945,WANG2022102082,FARSANE20001321,ghahfarokhi2020development,shams2020analytical} that cannot be easily solved using these conventional approaches. 
\begin{figure}
    \centering
    \includegraphics[width=0.5\linewidth]{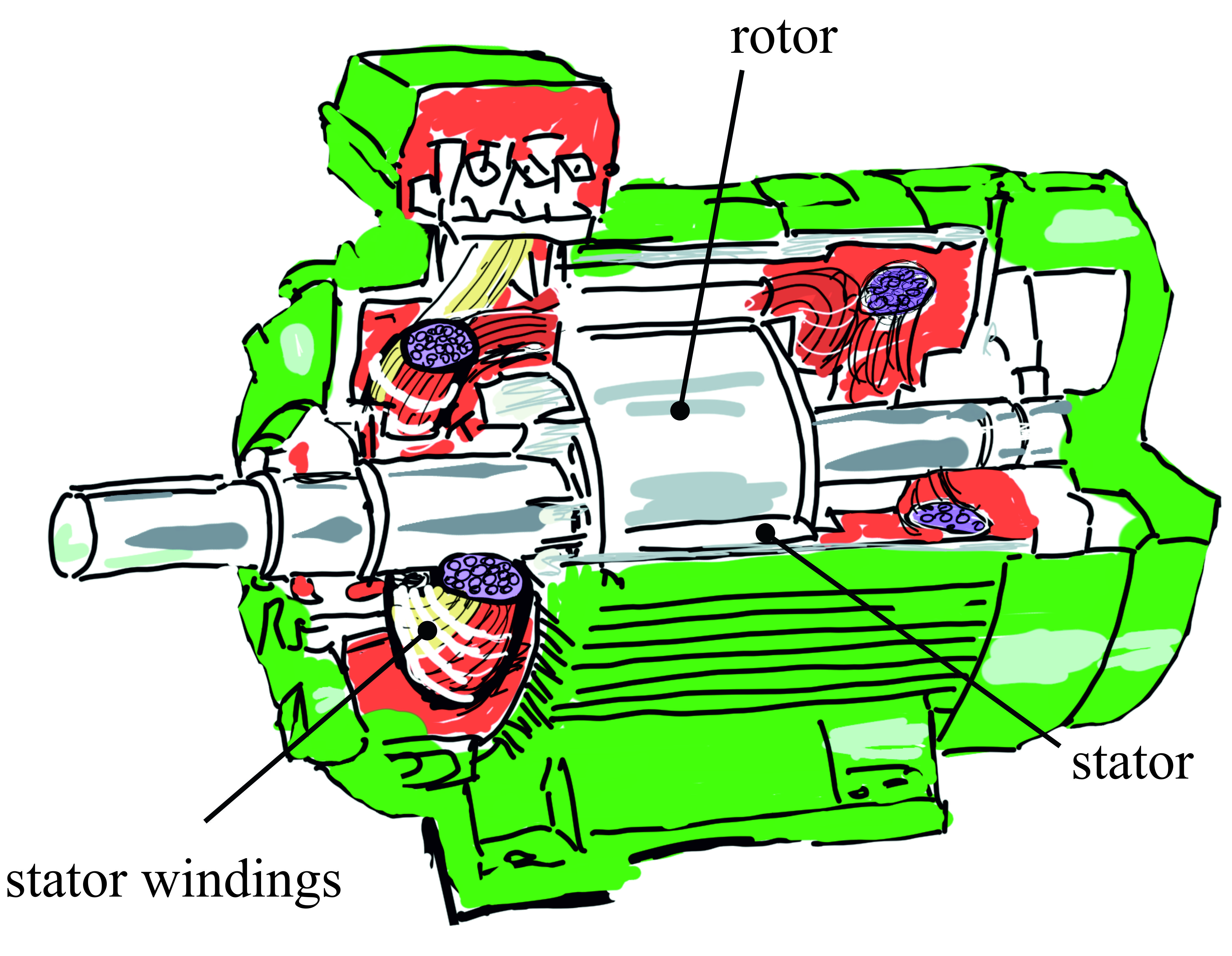}
    \caption{Schematic view of an electric machine.}
    \label{fig:EMSketch}
\end{figure}
Recently, spray cooling has started to be considered by engineers as a potentially very efficient technology for improving the thermal control of electric machines and other powerful electric and electronic systems \cite{chen2022application}. Experiments on the thermal management of electric machines  are usually focused on cooling the end winding \cite{8848870,10239016,zhang2021thermal,zhao2023parameter}, which is a relatively easily accessible and intensively heated part of the machine (Fig.~\ref{fig:EMSketch}). The application of multiple full-cone nozzles \cite{8848870} is recommended for achieving uniform cooling. In most cases, cooling the winding requires some effort to prevent the flow of the aerosol of the oil from dropping into the gap between the stator and rotor \cite{el2014advanced}. Nonetheless, some recent studies suggest using air/spray cooling, including winding and gap cooling of the electric machine \cite{dong2021performance}. 
The application of spray cooling to the thermal management of electric machines has also been investigated using \ac{cfd} computation, based on existing empirical models from the literature, in some cases improved according to experimental estimations of the heat flux \cite{5382944,10.1260/1757-482x.5.4.239,doi:10.1177/1757482X16653895}. In all these studies, several important parameters are missing. Foremost, the spray used for cooling is not properly characterized.  The average drop diameter (for example the Sauter mean diameter $D_{32}$), the average drop velocity and direction, the local mass flux density, and the temperature are the main  parameters influencing spray cooling. These parameters are often measured using phase Doppler instruments \cite{albrecht2013laser} or other optical methods \cite{tropea2011optical}. The heat flux is also determined by the thermal and rheological properties of the liquid as well as the surface and thermal properties of the substrate. This is why, for the sake of a more accurate model, it is important to understand the basic dynamics of drop, spray impact, and accompanied heat transfer, as occurring in spray cooling applications.
Spray impingement is a very effective means of removing heat from hot surfaces. Spray cooling processes are used in the field of metal production \cite{mudawar1994universal,hall1995experimental}, for cooling high-power electronic devices, data centers, and hybrid electric vehicles \cite{tilton1994liquid,yin2022spray}, for cooling laser diode scanners \cite{huddle2000thermal}, in aerospace applications \cite{liu2019applications}, and for cryogenic cooling of human tissue in medicine \cite{torres1999estimation} and other biomedical applications \cite{zhang2022advanced}. The advantage of spray cooling lies in the simplicity of its technical realization. The ability to direct the spray allows for covering a large cooling area. These advantages are complemented by the ability to provide relatively high heat fluxes, comparable with those produced by micro-channels, heat pipes, or other high-intensity heat removal technologies. Comprehensive literature reviews on spray cooling in various systems and the involved physical phenomena can be found in \cite{kim2007spray,liang2017review,Breitenbach2018aa}. 
Spray impact onto a solid substrate quickly forms a thin, fluctuating liquid wall film. This film is generated by the deposited liquid after drop impacts. The dynamics of the film are governed by viscous and capillary effects, as well as by flow disturbances caused by drop impacts, as shown in Fig.~\ref{fig:spray_impact}. This is why the phenomena of spray impact cannot be modeled as a simple superposition of drop impact events \cite{moreira2010advances}. This is also the reason why the modeling approaches describing the deposition mass ratio and the characteristics of spray cooling are mainly empirical. 
The outcome of a drop impact onto a wetted wall is determined by the material properties of the liquid, like density $\rho$, viscosity $\mu$ and surface tension $\sigma$, the thickness of the wall film $h_\mathrm{film}$, impact velocity $U_0$, impact angle  and initial drop diameter $D_0$. The corresponding dimensionless quantities include the Reynolds and the Weber numbers and dimensionless film thickness
\begin{equation}
\text{Re} = \frac{\rho D_0 U_0}{\mu},\quad \text{We} = \frac{\rho D_0 U_0^2}{\sigma},\quad \delta = \frac{h_\mathrm{film}}{D_0}.
\end{equation}
Typical outcomes of drop impact onto a wetted wall are shown in Fig.~\ref{fig:dropoutcomes}. They include drop rebound at very small Weber numbers. Rebound is governed by the stresses in a thin air gap between the impacting drop and the liquid film; deposition; formation of an uprising liquid sheet, formed as a result of the interaction of the flow in the lamella with the outer undisturbed film. At velocities exceeding the splashing threshold, drop impact leads to the generation of a secondary drop, splash. 
The splash is also the result of various hydrodynamic phenomena. Typical splash phenomena, i.e., prompt splash, corona splash, corona detachment and breakup of a central jet, are shown in Fig.\,\ref{fig:splashes}.
Among important results of drop impact studies \cite{yarin2017collision} are the splashing threshold \cite{cossali1997impact,mundo1995droplet}
\begin{equation}\label{eq:Knumber}
    K_\mathrm{splash}^{8/5} = 2100 + 5800 \delta^{1.44},\quad K\equiv \text{We}^{1/2} \text{Re}^{1/4},
\end{equation}
and the splashed mass ratio
\begin{equation}
    \eta_\mathrm{splashed\,mass} \approx 0.5 - 0.62 \exp[- K_\mathrm{primary}^{8/5}],
\end{equation}
where $K_\mathrm{primary}$ is the averaged $K - $ number, defined in (\ref{eq:Knumber}), computed using the averaged parameters of the primary spray before wall impact. The corresponding deposited mass ratio is therefore
\begin{equation}
    \eta_\mathrm{deposited\,mass} = 1-\eta_\mathrm{splashed\,mass}.
\end{equation}
since the mass lost due to evaporation during the short time of drop collision is relatively small. 
\begin{figure}
    \centering
    \includegraphics[width=0.9\linewidth]{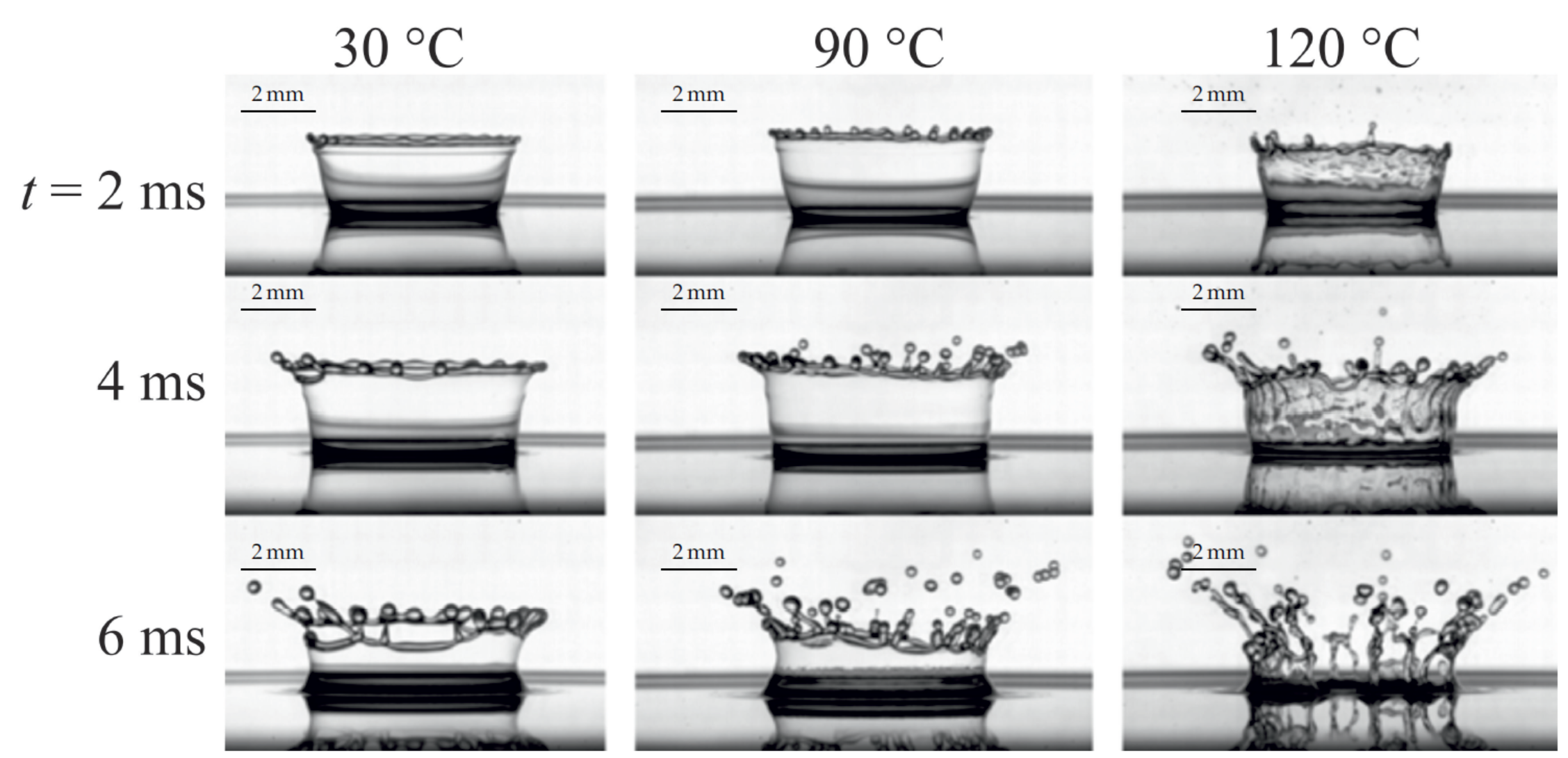}
    \caption{Effect of liquid temperature on the outcome of a Hexadecane drop impacting onto a Hexadecane wall film with $\delta =$ 0.1, $D_0$ = 2 mm and $U_0$ = 3 m/s.}
    \label{fig:droptemp}
\end{figure}
The outcome of drop impact can also be significantly influenced by the temperature, as shown in the example in Fig.~\ref{fig:droptemp}. This result can be explained by the dependence of viscosity and surface tension on temperature, as well as by the emergence of forces associated with thermal Marangoni effects or even liquid boiling. 
The model for heat flux associated with spray cooling depends on wall temperatures and thus on the impact of hydrodynamic and thermodynamic regimes. These regimes, at the highest wall temperatures, include drop rebound due to film boiling or drop thermal atomization. At intermediate temperatures, still significantly exceeding the boiling temperature, drop impact is accompanied by intensive nucleate boiling. At lower temperatures, below the boiling point, heat transfer occurs mainly by convection in the drop and in the liquid film, and by heat conduction in the substrate in the expanding thermal boundary layers.  
\begin{figure}
    \centering
    \includegraphics[width = 0.7 \linewidth]{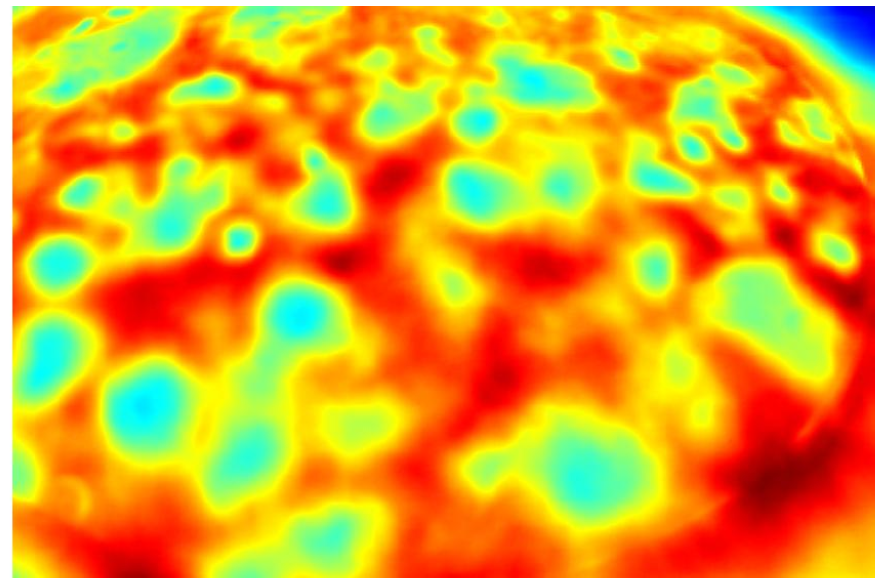}
    \caption{Spray-cooled target captured with an infrared camera. Red and yellow regions correspond to hot water film covering the target. Light blue and turquoise regions correspond to locations of recent impingement of cold droplets.}
    \label{fig:irspray}
\end{figure}
For relatively low temperatures, typical for the operating conditions of electric devices, the heat transfer is influenced by the expansion of the crater, formed in the film by drop impact. Heat transfer depends on the residual film thickness at the bottom of the crater, determined by the viscous stresses \cite{van2010dynamics,stumpf2022drop}, duration of the crater expansion, and receding and its maximum spreading diameter.\\
This paper focus on the simulation of the heat extraction from the winding overhang. The heat transfer coefficient is determined from an experiment (Sec.~\ref{Sec:5}), which included various volumetric fluxes, drop velocities and diameters. In contrary to the \ac{cfd} simulations in \cite{5382944,10.1260/1757-482x.5.4.239,doi:10.1177/1757482X16653895}, the \ac{fe} method is applied, with the assumption of a uniformly distributed spray \cite{8848870}. Additionally, a quasi-3D approach is exploited, which is successfully used in other physical simulations as in \cite{DAngelo_2020, WEN2019431,https://doi.org/10.1049/iet-epa.2015.0491}, and avoids an unaffordable \ac{3d} simulation.

\section{Thermal Model}
\subsection{Thermal machine model}
During electric machine operation, there are several sources of heat. In the stator and rotor windings and bars, Joule losses are generated. In the yoke parts, eddy-current and hysteresis losses are generated. At the interface between moving parts friction losses occur. A stator winding consists of wires separated by insulation. A part is mounted in the stator slots. The overhang winding parts stick out at the front and back sides of the machine. The interturn insulation is considered one of the weakest parts of an electric machine \cite{Driendl_2022aa} and is, therefore, a common reason for machine failures. This paper will focus on the cooling of the stator windings and thus consider the Joule losses as the major heat source. The inside parts of the stator windings are hot spots. Commonly, the heat is extracted from the  overhang winding parts by air cooling. In this paper, spray cooling of the overhang winding parts is studied as an alternative.\\
As an example machine model, a squirrel-cage induction machine located at the Electric Drives and Power Electronic Systems Institute of TU Graz is considered \cite{Eickhoff_2020aa}. Its thermal properties have been investigated in \cite{bergfried2023aa, 10700088}. The machine is designed for a nominal power of 3.7\,kW at a nominal speed of 1430 rpm. Moreover, it is rated for air-cooling but is surrounded by a cooling jacket for measurement purposes.\\
\subsection{Heat conduction model}\label{sec:thermal_model}
In the stationary case, the heat equation \cite{Chemieingenieurwesen_2010aa} is 
\begin{alignat}{2}
    -\nabla\cdot\left(\lambda\nabla\vartheta\right)
    &=p, &\qquad\text{in }\mathcal{V}\subset \mathds{R}^3 ;\\
    \vartheta &=\vartheta_{\text{iso}}, &\qquad\text{at }\mathcal{S}_{\text{iso}};\\
   \vec{n}\cdot\vec{q} &=0, &\qquad\text{at }\mathcal{S}_{\text{adia}} ;\\
    \label{eq:impBC}\vec{n}\cdot\vec{q} &=g_{\text{imp}}(\vartheta), &\qquad\text{at }\mathcal{S}_{\text{imp}},
\end{alignat}
where $\vartheta(\vec{r})$ is the temperature, $\vec{q}(\vec{r}=-\lambda\nabla\vartheta$ is the heat flux density, $\lambda(\vec{r},\vartheta(\vec{r}))$ is the thermal conductivity, $p(\vec{r})$ is the internal heat source density, in the case of a stator winding, the Joule loss density. $V$ is the computational domain, $\partial V = S_{\text{iso}} \cup S_{\text{adia}} \cup S_{\text{imp}}$ its surface divided into an isothermal, adiabatic and impedance part, $\Vec{n}(\Vec{r})$ is the unit normal vector at $\partial V$ and $\Vec{r}= (x, y, z)$ is the coordinate system, where z is direction along the axis of the machine. \\
The isothermal \ac{bc} enforces a prescribed temperature distribution $\vartheta_{\text{iso}}$ at $S_{\text{iso}}$. The adiabatic \ac{bc} prescribes the heat flux density, here 0. The impedance \ac{bc} imposes a relation $g_{\text{imp}}$ between heat flux density and the temperature at the boundary.
\subsection{Simplification of the thermal model}\label{sec:simp_thermal}
To reduce model complexity and computational effort, the following assumptions are considered:
\begin{itemize}
    \item The problem is stationary: This assumes a constant internal and external heat source density. As a consequence, the heat generated in the domain equals the heat flux through the surface.
    \item The radial heat flux within the machine is negligible: This implies that the generated heat within the stator winding flows towards the overhang winding, and is there removed by spray cooling.
    \item The properties of the spray cooling are constant, i.e., the liquid film has a constant thickness and a constant temperature.
    \item The central part of the geometry is invariant in z-direction. The problem is formulated for the 2D cross-section of the machine. Nevertheless, some fields feature are axial dependence, which will be taken into account by a quasi-3D formulation.
    \item To model the spray cooling effects, the machine model gets reduced to a stator winding.
    %\item \cb{The ratio between the length of the wire within the slot and in the end-winding part is 3/5 to 2/5.}
\end{itemize}
With these assumptions, the thermal model gets reduced to
\begin{alignat}{2}
     -\nabla\cdot\left(\lambda\nabla\vartheta\right)
    &=p, &\qquad\text{in }\Omega_k ;\label{eq:heat_eq_2d}\\
    -\lambda \nabla \vartheta \cdot\vec{n} &= 0, &\qquad\text{at }\Gamma_{\text{adia}};\\
    \lambda \nabla \vartheta\cdot\vec{n} + h(\vartheta-\vartheta_0) &= 0, &\qquad\text{at }\Gamma_{\text{imp}},
\end{alignat}
with 
\begin{alignat*}{2}
    \Omega_k &:= \{\vec{r}=(x,y,z)\in\mathcal{V}\,|\,z=f(s)\} & \text{for }k\leq |N|\in \mathds{Z};\\
    \Gamma_{\text{adia}} &:= \{\vec{r}=(x,y,z)\in \partial\Omega_k\cap\mathcal{S}_{\text{adia}}\} & \text{for }k<|\frac{3}{5}N| \in \mathds{Z};\\
    \Gamma_{\text{imp}} &:= \{\vec{r}=(x,y,z)\in \partial\Omega_k\cap\mathcal{S}_{\text{imp}}\} & \text{for }k>|\frac{3}{5}N| \in \mathds{Z};
    %\Gamma_{\text{imp}} &:= \{\vec{r}=(x,y,z)\in \{ \Omega_0,\Omega_N\}\,\widehat{=}\, \mathcal{S}_{\text{imp}}\}\\
    %\Gamma_{\text{adia}} &:= \{\vec{r}=(x,y,z)\in \partial\Omega_k\cap\mathcal{S}_{\text{adia}}\} & \text{for }k<|N| \in \mathds{Z},
\end{alignat*}
where $\Omega_k$ are discrete cross-sections of $\mathcal{V}$, $\Gamma_{\text{imp}}$ and $\Gamma_{\text{adia}}$ are the boundaries. %, and $\Delta z$ and $2N+1$ are the slice-size and number of slices, respectively, of the quasi-3D approach.
$h$ is the heat transfer coefficient at $\Gamma_{\text{imp}}$ modeling the spray cooling at the overhang winding surface.
The field problem is solved with the \ac{fe} solver \texttt{Pyrit} \cite{Bundschuh_2023ab} and validated using \texttt{COMSOL Multiphysics}$^\text{\textregistered}$ \cite{Comsol}. 
The heat source density equals the Joule losses \cite{meschede2015gerthsen}
\begin{equation}
    p=\frac{J^2}{\sigma_{\text{cu}}}, \label{eq:Joule_loss}
\end{equation}
where $J$ is the current density and $\sigma_{\text{cu}}$ is the electric conductivity of copper.
\section{Discretization}
The assumption of a translatory invariant geometry in the winding direction allows for a \ac{2d} cross-sectional \ac{fe} model which is extruded in the winding direction using the quasi-3D approach.
\subsection{Finite-element}
The \ac{2d} cross-section is meshed with triangles. The temperature is approximated as
\begin{equation}
    \vartheta \approx \sum_{j=1}^N u_j N_j \label{eq:discretisation_2d}
\end{equation}
where $N_j(x,y)$ are \ac{2d} first-order nodal shape functions and $u_j$ the degrees of freedom.\\
Using the Ritz-Galerkin approach \cite{Fletcher.1984}, the thermal model \eqref{eq:heat_eq_2d} leads to the system of equations
\begin{equation}
(\mathbf{K}_{\lambda} + \mathbf{M_{h}})\mathbf{u}=\mathbf{p}+\mathbf{q}\label{eq:sys_of_eq}
\end{equation}
with the stiffness matrix $\mathbf{K}_{\lambda}$, the mass matrix $\mathbf{M_{h}}$ and the internal load vector $\mathbf{p}$ and external load vector $\mathbf{q}$ expressed.
\begin{alignat}{1}
    K_{\lambda,ij} &= \int_{\Omega} \lambda \nabla N_j \cdot \nabla N_i\,\text{d}\Omega;\\
    M_{h,ij} &= \int_{\Omega} h N_j N_i\,\text{d}\Omega;\\
    p_i&=\int_\Omega p N_i \, \text{d}\Omega;\label{eq:p_i}\\
    q_i&=\int_\Gamma h \vartheta_0 N_i \,\text{d}\Gamma.\label{eq:q_i}
\end{alignat}
%Note that $h_{\text{spray}}$ is defined as 
%\begin{equation}
%    h_{\text{spray}}(\vec{r})=
%    \begin{cases}
%        \frac{\text{Nu} \lambda_{\text{film}}}{h_\text{film}} &\qquad \text{if}\enspace %\vec{r} \in \Gamma_{\text{imp}},\\
%        0 &\qquad \text{if}\enspace \vec{r} \not\in \Gamma_{\text{imp}},
%    \end{cases}\label{eq:heat_transfer_coef}
%\end{equation}
%where Nu is the Nusselt number. This means that there is only a heat transfer at the %impedance boundary. 
\subsection{Quasi-3D approach}\label{sec:Quasi_3D}
The winding direction is discretized by 
\begin{equation}
    z=f(s),
\end{equation}
where $f(s)$ is a Frenet-Serret mapping \cite{9233257}. 
%\begin{figure}[h!]
%    \centering
%    \includegraphics[width=0.8\linewidth]{example-image}
%    %\resizebox{0.8\linewidth}{!}{
%    \input{figs/Shapefunction}%}
%    \caption{To be deleted. Placeholder that the Figure numbers still match the comments from Anouar}
%    \label{fig:shapefunction}
%\end{figure}%
For this, \ac{1d} quadratic lagrange shape functions $N_{\text{k}}^{\text{1D}}(z)$ are introduced as 
\begin{align}
    N_1(\xi) &= \frac{1}{2}\xi(\xi-1),\\
    N_2(\xi) &= (1-\xi^2),\\
    N_3(\xi) &= \frac{1}{2}\xi(\xi+1).
\end{align}
Here $\xi$ is a reference \ac{1d} element with $\xi\in[-1,1]$.
Therefore, the approximation in \eqref{eq:discretisation_2d} becomes the quasi-3D formulation 
\begin{equation}
    \vartheta(u,v,s) \approx \sum_{j=1}^N \sum_{k=1}^N u_{jk} N_j^{\text{2D}}(u,v) N_k^{\text{1D}}(s). 
\end{equation}
With this approximation of the temperature, the stiffness matrix $\mathbf{K}^{\text{3D}}$ is derived as \cite{DAngelo_2020}
\begin{alignat}{1}
    K_{\lambda,ijkl} &= \int_{\Omega} \lambda \nabla (N_j^{\text{2D}} N_k^{\text{1D}}) \cdot \nabla (N_i^{\text{2D}} N_l^{\text{1D}})\,\text{d}\Omega;\notag\\
    &=\underbrace{\int_{\Omega} \lambda \nabla N_j^{\text{2D}} \cdot \nabla N_i^{\text{2D}} N_k^{\text{1D}} N_l^{\text{1D}}\,\text{d}\Omega}_{M^{\text{1D}}_{kl}\otimes K^{\text{2D}}_{ij}} \\
    &+\underbrace{\int_{\Omega}\lambda N_j^{\text{2D}} N_i^{\text{2D}} \nabla N_k^{\text{1D}} \cdot \nabla N_l^{\text{1D}}\,\text{d}\Omega}_{K^{\text{1D}}_{kl}\otimes M^{\text{2D}}_{ij}} .\notag
\end{alignat}
The \ac{3d} stiffness matrix $\mathbf{K}^{\text{3D}}$ is constructed using \ac{1d}- and \ac{2d}-stiffness, $K$ and mass matrix, $M$, which are multiplied using the Kronecker product denoted as $\otimes$.
A similar adjustment is necessary for the \ac{3d} mass matrix $\mathbf{M^{\text{3D}}}$ and the loads $\mathbf{q}$ and $\mathbf{p}$
\begin{align}
    M_{h,ijkl} &= \underbrace{\int_\Omega h N_j^{\text{2D}} N_i^{\text{2D}} N_k^{\text{1D}} N_l^{\text{1D}}\,\text{d}\Omega}_{M^{\text{1D}}_{kl}\otimes M^{\text{2D}}_{ij}},\\
    q_{il}^{\text{3D}} &= \underbrace{\int_\Omega h\vartheta_0 N_i N_l \,\text{d}\Omega}_{q^{\text{1D}}_i\otimes\, q^{\text{2D}}_l},\\
    p_{il}^{\text{3D}} &= \underbrace{\int_\Omega p N_i N_l \,\text{d}\Omega}_{p^{\text{1D}}_i\otimes\, p^{\text{2D}}_l}.
\end{align}
\section{Quasi-3D FE Model of a Stator Winding with Spray Cooling}
\subsection{Geometry}
\begin{figure}[h!]
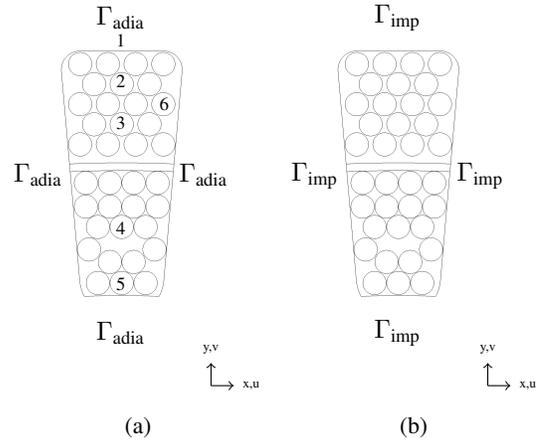

    \centering
    \begin{subfigure}{0.4\linewidth}
    %\resizebox{1\linewidth}{!}{
    \input{figs/tikz_slot_model_1}%}
    \caption{}
    \label{fig:tikz_slot_model_1}
    \end{subfigure}
    \begin{subfigure}{0.4\linewidth}
    \centering
    %\resizebox{1\linewidth}{!}{
    \input{figs/tikz_slot_model_2}%}
    \caption{}
    \label{fig:tikz_slot_model_2}
    \end{subfigure}
    \caption{(a): \ac{2d} cross-section of the winding within the stator with the corresponding boundary conditions. The numbers indicate the points that are used for temperature evaluation. (b): \ac{2d} cross-section of the overhang-winding outside the stator. Only the overhang winding is getting spray-cooled.}
    \label{fig:Tikz_slot}
\end{figure}

%\cb{To model the spray cooling effects, the machine model gets reduced to a stator winding. 
The winding model is displayed in Fig.~\ref{fig:Tikz_slot} as two cross-sections with different boundary conditions and in Fig.~\ref{fig:model_3D} as a \ac{3d} sketch.
Due to symmetry, it is sufficient to consider only a part of a single stator winding extending between the center cross-section of the machine (z=0 in Fig.~\ref{fig:model_endwinding_curved}) to the outermost point of the overhang (point p in Fig.~\ref{fig:model_endwinding_curved}). The curved geometry is projected onto a straight (Fig.~\ref{fig:model_endwinding}) by the coordinate transformation (u,v,s) $\rightarrow$ (x,y,z) = (f(u,v,s), g(u,v,s), h(u,v,s)). To simplify the setting, u and v are chosen to match x and y in the center cross-sectional plane, and s is chosen to be the length measured along the center line of the winding. Moreover, the curvature torsion of the winding is neglected. These assumptions are acceptable for the thermal problem considered here but would not work for a magnetic model. In this simplified setting, a quasi-3D FE formulation of section~\ref{sec:Quasi_3D} can be applied without distinguishing between (u,v,s) and (x,y,z). \\
The cross-section of the quasi-3D model is depicted in Fig.~\ref{fig:Tikz_slot}. The area of the cross-section is 107.7\,mm$^2$ with two-layer wire windings immersed in insulating epoxy resin. Each layer consists of 18 copper wires, resulting in a copper fill factor of 59\,\%. Other geometrical and thermal properties of the winding, used in the computations, are listed in Table~\ref{tab:parameter}.
\subsection{Boundary condition}
\begin{figure}[h!]
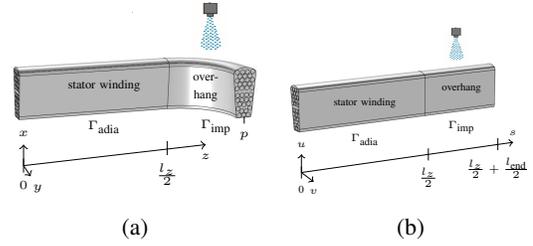

    \centering
    \begin{subfigure}{0.4\linewidth}
    \resizebox{1\linewidth}{!}{
    \input{figs/Model_Comsol_Curved}}
    \caption{}
    \label{fig:model_endwinding_curved}
    \end{subfigure}
    \begin{subfigure}{0.4\linewidth}
    \centering
    \resizebox{1\linewidth}{!}{
    \input{figs/Model_Comsol}}
    \caption{}
    \label{fig:model_endwinding}
    \end{subfigure}
    \caption{Simulation model of stator winding and overhang in different coordinate system.}
    \label{fig:model_3D}
\end{figure}

Due to symmetry, there is no heat flux through the center and outermost cross-sectional planes (s=0 and s=$\frac{l_z}{2}+\frac{l_{end}}{2}$). Hence, adiabatic boundary conditions $-\lambda\frac{\partial \vartheta}{\partial s}=0$ are applied there.\\
Within the stator yoke, the winding is surrounded by good electric insulation, which also provides good thermal insulation. As a consequence, the heat flux towards the stator yoke can be neglected. Hence, an adiabatic BC $-\lambda\frac{\partial \vartheta}{\partial s}=0$ is applied (Fig.~\ref{fig:model_endwinding_curved}). \\
The winding overhang is subjected to spray cooling. It is assumed that the spray is homogeneously applied to the winding surfaces. The inhomogeneous temperature distribution, however, will cause inhomogeneous heat removal. The spray-cooling effect is modeled as an impedance BC $\Vec{n}\cdot\Vec{q}=g_{\text{imp}}(\vartheta)=h_{\text{spray}}(\vartheta-\vartheta_0)$ with $h_{\text{spray}} = 22485\,$W/Km$^2$ and $\vartheta_0 = 293\,$K, as determined by the spray-cooling experiment (Section IV, Table II).
\begin{table}
    \centering
    \caption{Parameters of the winding model}
    \begin{tabular}{l c r l}
    \toprule
       \textbf{ Name} &\textbf{ Parameter} &\textbf{ Value} & \textbf{Unit}\\
         \midrule
        Length within stator & $l_\text{s}$ & 100 & mm\\
        Length overhang & $l_{\text{ew}}$ & 66.6 & mm\\
        Current density & $J$ & 10 & A/mm$^2$\\
        Electric conductivity copper& $\sigma_{\text{cu}}$ & 60 & MS/m\\ %59.98
        Thermal conductivity copper & $\lambda_{\text{cu}}$& 400 & W/Km\\
        Thermal conductivity insulation & 
        $\lambda_{\text{ins}}$ & 0.7 & W/Km\\
        Thermal conductivity of water & $\lambda_{\text{film}}$& 0.6 & W/Km\\ %0.598
                Heat source density & $p$ & 1.7& MW/m$^3$ \\ %1667222
                \bottomrule
    \end{tabular}
    \label{tab:parameter}
\end{table}%
\section{Experimental Characterization of the Spray Cooling}\label{Sec:5}
\begin{figure}
    \centering
    \includegraphics[width = 0.7\linewidth]{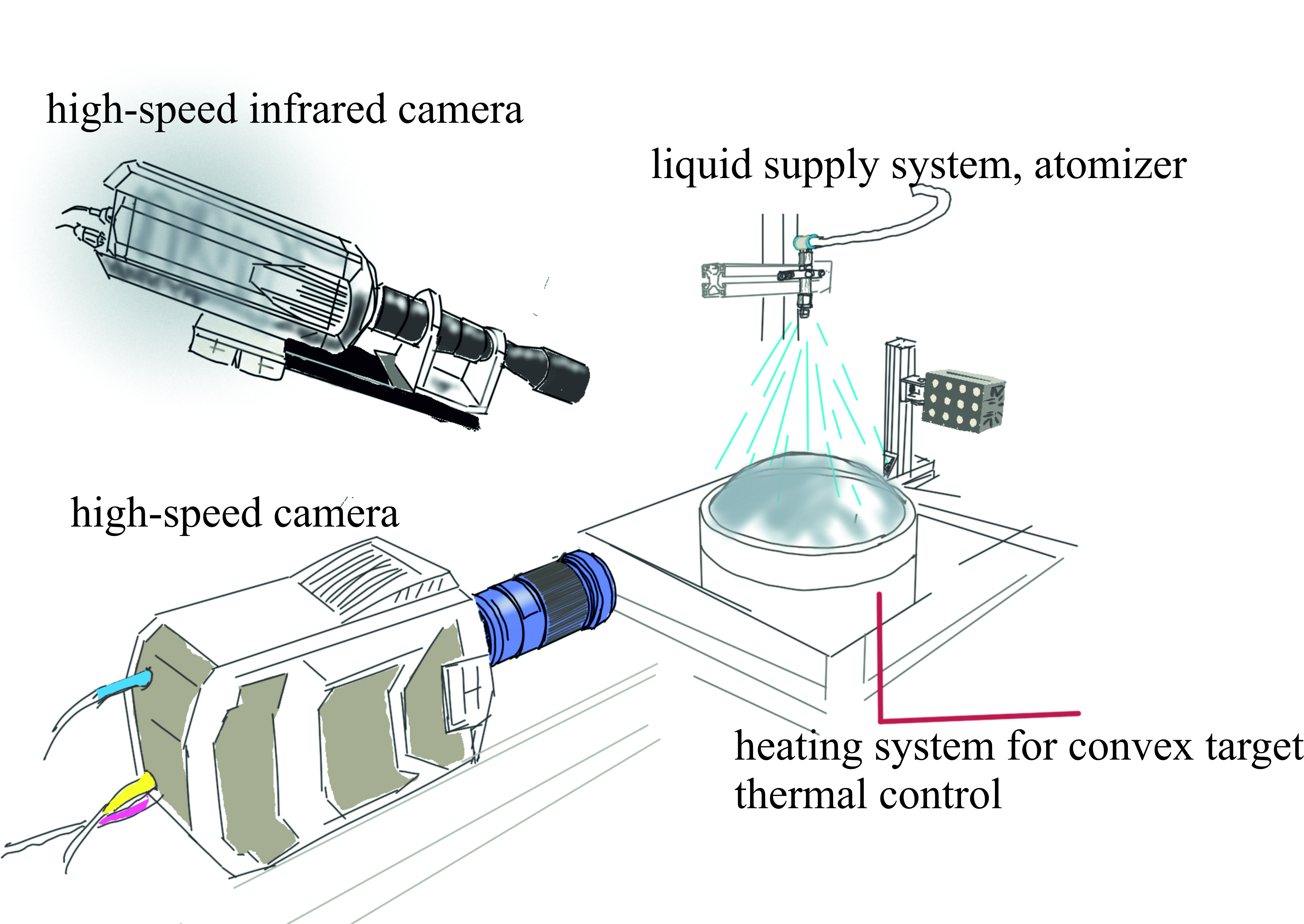}
    \caption{Sketch of the experimental setup used to investigate spray cooling. More details can be found in  \cite{kyriopoulos2010gravity}.}
    \label{fig:spray_cooling_setup}
\end{figure}
Figure \ref{fig:spray_cooling_setup} depicts an experimental setup \cite{kyriopoulos2010gravity} for studying spray impact onto a hot target and measuring the heat flux for various spray parameters: volumetric flux, drop velocity, and diameter. The setup consists of the following major sub-systems: a liquid supply system and atomizer for water spray generation, a convex target with a heating and thermal control system, and the observation system, which includes a high-speed video system with 16000\,fps (FASTCAM SA1, Photron) and high-speed infrared system with 100 frames at 244\,Hz (Phoenix, Indigo Systems). 
The spray has been  characterized using a phase Doppler instrument. The heat flux in the target was measured using an array of thermocouples. The metal target of convex shape included a carefully calibrated thin thermal resistance layer, parallel to the interface. This allows a significant increase in the temperature jump in the thermocouple measurements and thus improves the precision of the heat flux measurements.   
The flow on the target surface is characterized by a high-speed infrared camera (see Fig.~\ref{fig:irspray}) and a high-speed camera (see e.g. in Fig.~\ref{fig:spray_impact}). The average film thickness at the generatrix of a convex target surface was measured using a high-resolution video camera with 36000\,fps (HighSpeedStar 5, LaVision). 
The experimental setup has been developed for investigating the effect of gravity acceleration on the heat flux associated with spray impact \cite{gambaryan2007gravity}. In this study, however, only results obtained under normal terrestrial conditions are relevant. The Sauter mean diameter $D_{32}$ of the drops in the spray was varied from 390 to 890 $\mu$m, the average impact velocity was chosen in the range between 5.6 and 14 m/s, the  volumetric flow rate of the liquid was specified between 0.25 and 0.65 l/min. The spray angle is 30$^\circ$ and the distance to the target is 70\,mm. Therefore, the mass flux $\dot m$ can be roughly estimated from the mass balance and geometrical considerations, to be between 3.8 and 9.8\,kg/m$^2$ s. 
Interestingly, for these relatively wide ranges of the main spray parameters, the variations of the measured heat transfer coefficients $h_\mathrm{spray}$ are rather weak: from $2.2\times 10^4$ to $2.3\times 10^4$ W/m$^2$K. 
Hence, the dependencies of the spray-cooling effect on the spray-cooling parameters are neglected in the \ac{fe} simulations reported below. A representative setting of the spray-cooling has been selected, for which the parameters, among the heat transfer coefficient $h_{\text{spray}}$ and the environmental temperature used in the simulations, are listed in Table~\ref{tab:parameter_calculated}. 
\begin{table}
    \centering
    \caption{Parameters used for computations}
    \begin{tabular}{l c r l}
    \toprule
        \textbf{Name} & \textbf{Parameter} & \textbf{Value} & \textbf{Unit}\\
         \midrule
         Mass flow rate & $\dot M$ & 0.35 & kg/min\\
         Estimated mass flux &$\dot m$ & 5.3 &kg/m$^2$s\\
         Average spray velocity & $U_\mathrm{spray}$ & 7.77 & m/s\\
         Sauter mean diameter & $D_{32}$& 497 & $\mu$m\\
         Spray temperature & $\vartheta_0$ & 293 & K \\
    Film thickness & $H_\mathrm{film}$ & 91 & $\mu$m \\
        Heat transfer coefficient & $h_\mathrm{spray}$ & 22.5 & kW/Km$^2$\\ %22485
        \bottomrule
    \end{tabular}
    \label{tab:parameter_calculated}
\end{table}%
\section{Results and Discussion}
\subsection{Existing machine}
\begin{figure}
    \centering
    \includegraphics[width = 0.6\linewidth]{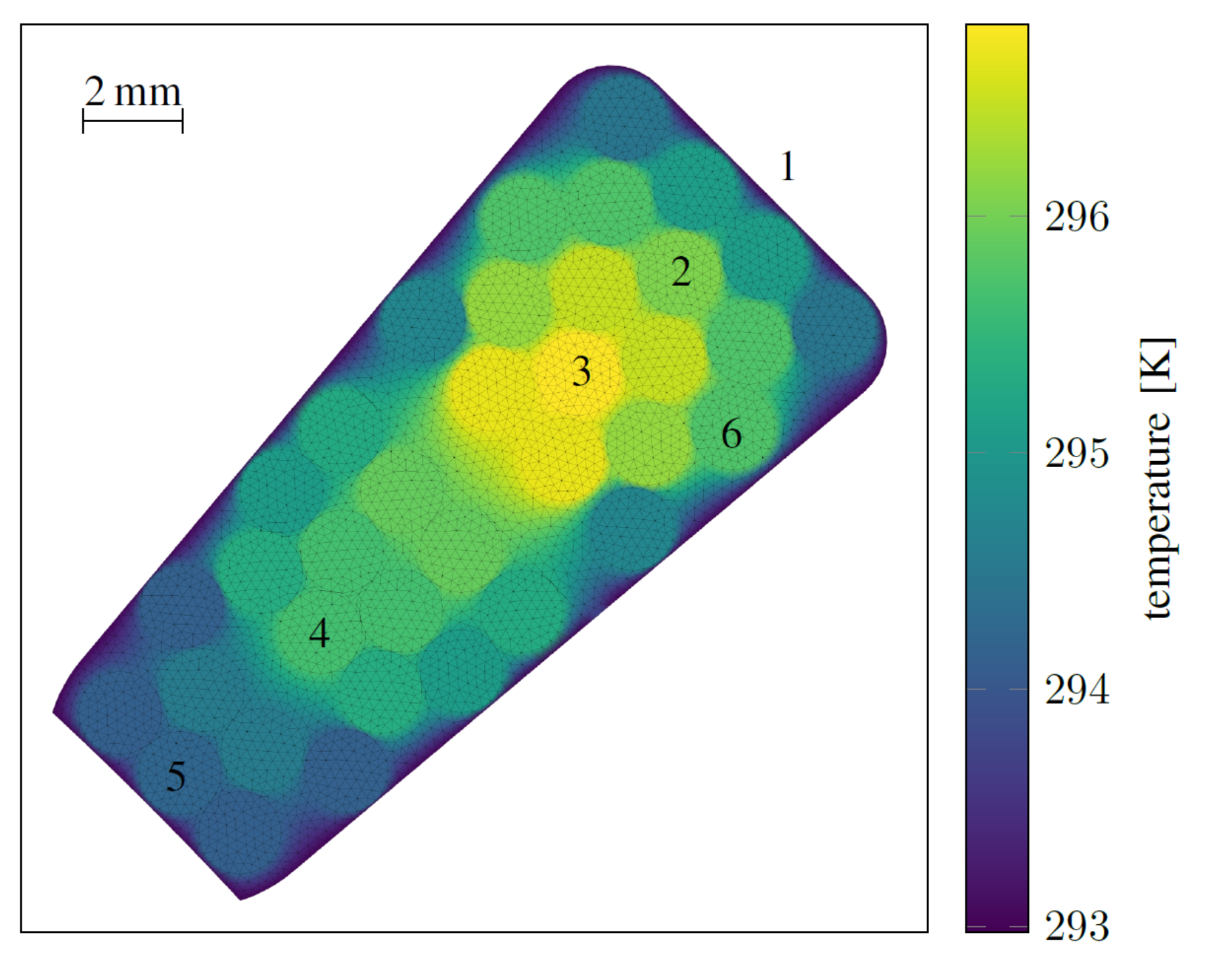}
    \caption{Temperature in the cross-section at the most outside point (s = $\frac{l_z}{2}+\frac{l_\text{end}}{2}$). The distribution is mainly influenced, by the spray-cooling boundary condition.}
    \label{fig:cross_section_endwinding}
\end{figure}

\begin{figure}[h!]
    \centering
    \includegraphics[width = 0.8\linewidth]{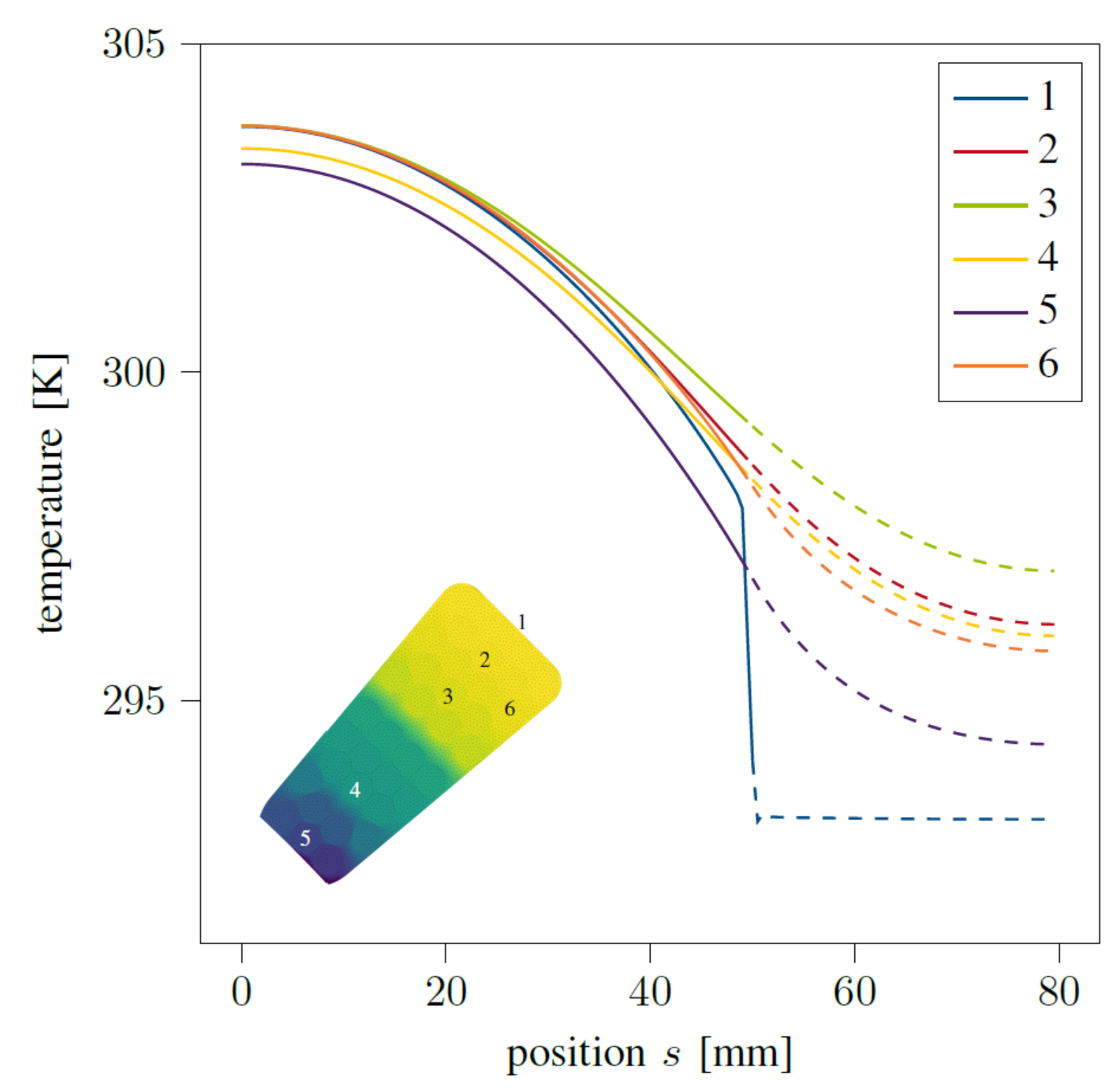}
    %\resizebox{1\linewidth}{!}{
    %\input{figs/temp_along_s}}
    %\input{figs/Temperature_along_axis}}
    \caption{Temperature along the winding. The solid line represents the temperature in the winding inside the slot and the dashed line represents the temperature in the overhang part of the winding. To get the results for an entire winding, the result should be mirrored several times at the left and right positions.} %The positions the points of within the winding are shown in Fig.~\ref{fig:tikz_slot_model_1}, \ref{fig:cross_section_slot} and \ref{fig:cross_section_endwinding}.}
    \label{fig:temp_axis}
\end{figure}

%Figures \ref{fig:cross_section_slot} and \ref{fig:cross_section_endwinding} display the temperature distribution in the center and the most outside cross-sections, respectively. 
In the central cross-section, the maximal temperature difference is 0.5\,K. Moreover, the individual wire cannot be recognized because the temperature distribution is almost homogeneous (see Fig.~\ref{fig:temp_axis} lower left corner). In the most outside cross-section, displayed in Fig.~\ref{fig:cross_section_endwinding}, the temperature difference is 3\,K from the winding center to the surface. Moreover, each wire is visible, as a temperature gradient from the center to the surface exists.
%\begin{figure}[h!]
%    \centering
%    \includegraphics[width=0.8\linewidth]{example-image}
    %\resizebox{0.8\linewidth}{!}{
%    \input{figs/Shapefunction}%}
%    \caption{To be deleted. Placeholder that the Figure numbers still match the comments from Anouar}
%    \label{fig:shapefunction}
%\end{figure}%
%\begin{figure}[h!]
%    \centering
    %\resizebox{0.8\linewidth}{!}{
    %\input{figs/Temperature_distribution_Slot}}
    %\caption{Temperature in the center the cross-section of the machine (z = s = 0). The distribution is mainly influenced by the adiabatic boundary conditions.}
    %\label{fig:cross_section_slot}
%\end{figure}%
%\begin{figure}[h!]
%    \centering
%    \resizebox{0.65\linewidth}{!}{
%    \input{figs/Temperature_distribution_endwinding}}
%    \caption{Temperature in the cross-section at the most outside point (s = $\frac{l_z}{2}+\frac{l_\text{end}}{2}$). The distribution is mainly influenced, by the spray-cooling boundary condition.}
%    \label{fig:cross_section_endwinding}
%\end{figure}%
The resulting temperature distribution along the winding direction is shown in Fig.~\ref{fig:temp_axis} for the temperature at selected points. The temperature distribution along the wire winding can be retrieved by mirroring the curves in Fig.~\ref{fig:temp_axis} left and right. Therefore, the position (s = 0\,mm) is in the middle of the stator, the position (s = $\frac{l_z}{2}$ = 50\,mm) is the end of the stator and the beginning of the overhang, and the position (s = $\frac{l_z}{2}+ \frac{l_{\text{end}}}{1}$ = 80\,mm) is the middle of the overhang. At the center cross-section (at 0\,mm), the temperature in the center and at the surface are almost identical. This is attributed to the good thermal insulation between the winding and yoke. A temperature drop of 6\,K is located at the end of the stator and at the beginning of the overhang. From the center cross-section, the temperature decreases continuously in the winding direction. The highest temperature is in the lower layer of the winding, with 304\,K within the slot and 296\,K in the overhang. As this is the point with the largest distance to the surface, the cooling effect of the spray is the lowest there. The lowest temperature is found at the surface with the spray impact. It is only marginally larger than the spray temperature. Even within the center cross-section, where the winding is not cooled, the temperature of the wires remains at a moderate level. This is because the investigated machine is rated for air cooling.
\subsection{Machine with increased power density}
\begin{figure}[h!]
    \centering
    \includegraphics[width = 0.8\linewidth]{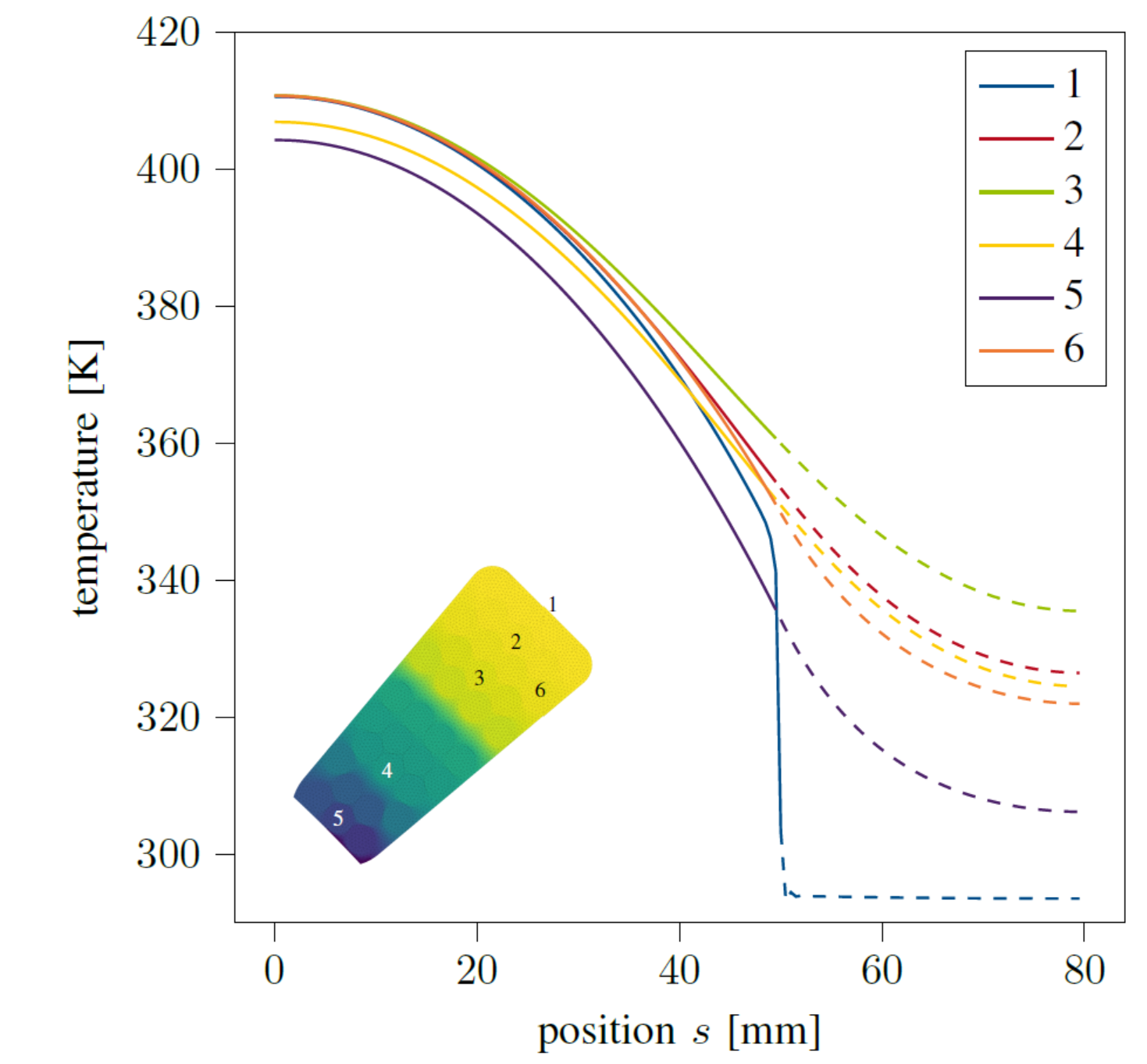}
    %\resizebox{1\linewidth}{!}{
    %\input{figs/temp_along_s_3}}
    \caption{Temperature along the winding for a 10 times higher power density.} %The positions of the points within the winding are shown in Fig.~\ref{fig:tikz_slot_model_1} and \ref{fig:cross_section_endwinding}.}
    \label{fig:temp_axis_5}
\end{figure}
\begin{figure}
    \begin{subfigure}{0.45\linewidth}
    \includegraphics[width=0.9\linewidth]{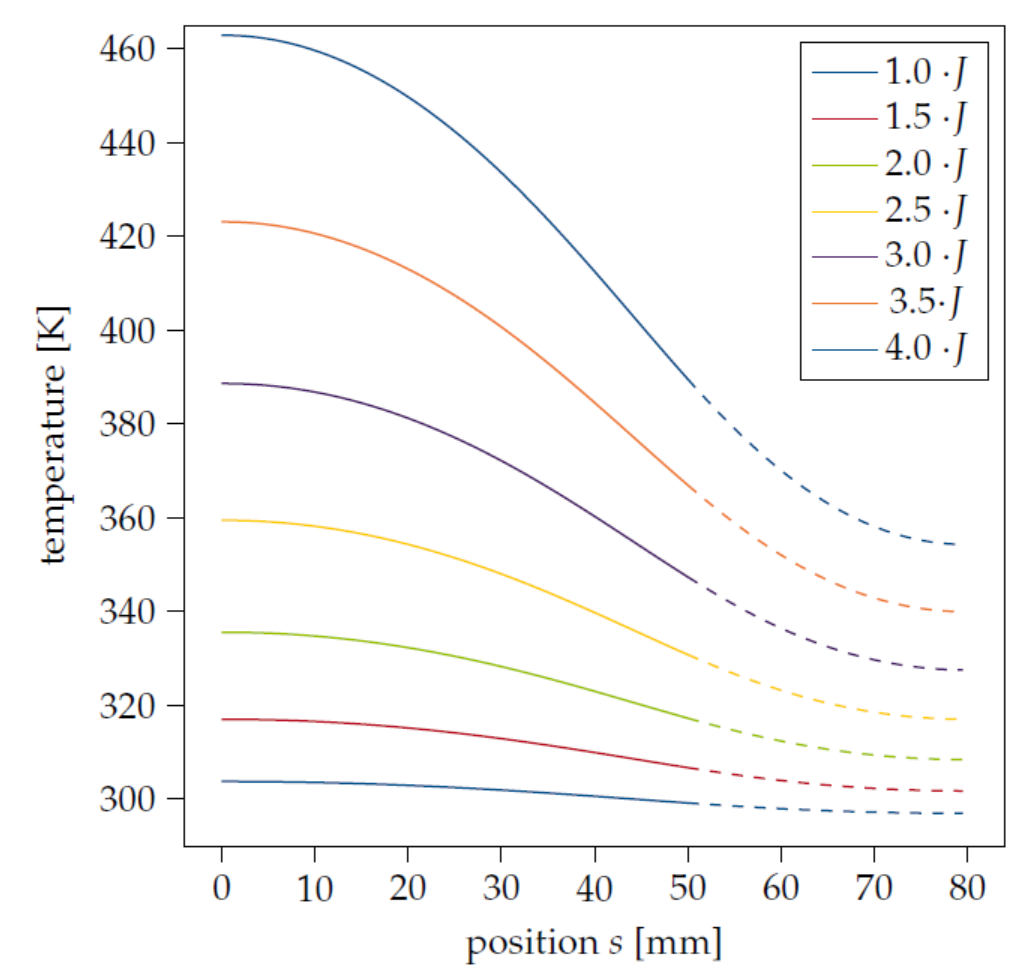}
    %\resizebox{0.9\linewidth}{!}{
    %\input{cooling_different_joule}}
    \caption{Variation of the current density. The current density is $J=10\,$A/mm$^2$.}
    \label{fig:investigation_J}
    \end{subfigure}
    \begin{subfigure}{0.45\linewidth}
    \includegraphics[width=0.9\linewidth]{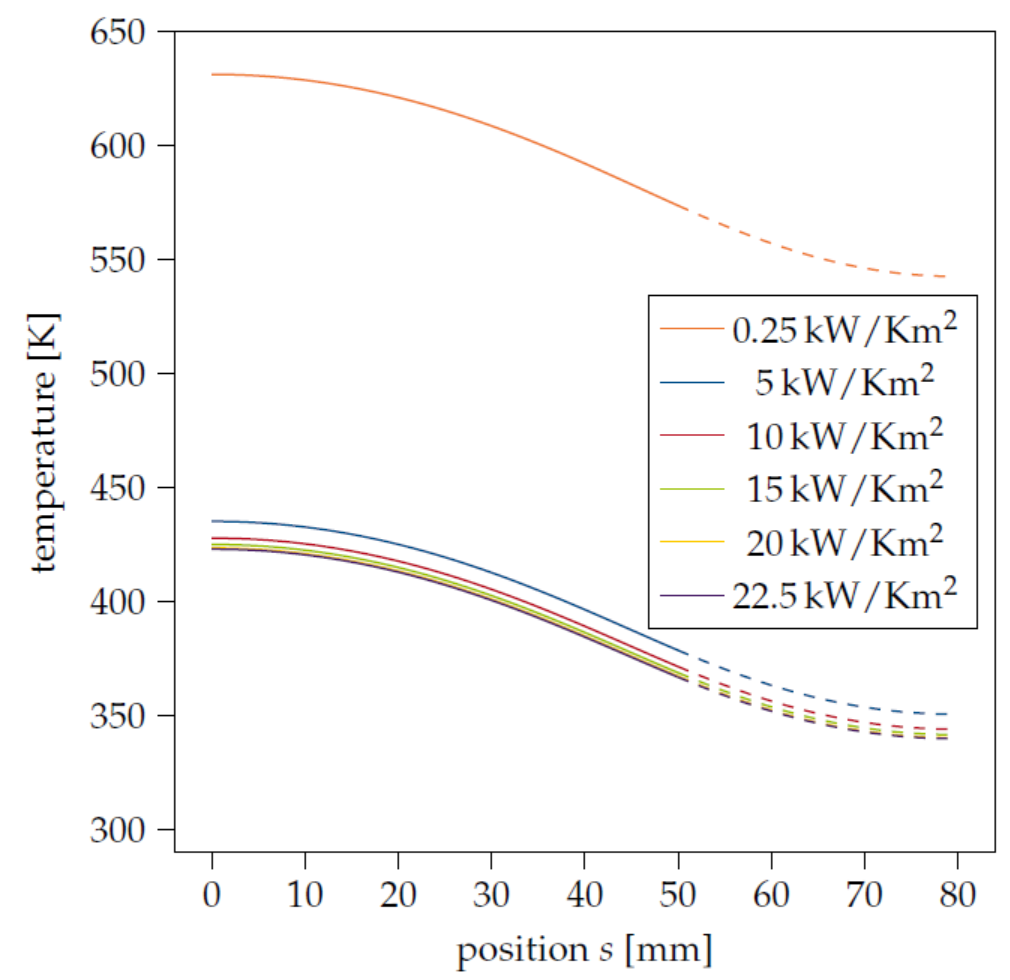}
    %\resizebox{0.9\linewidth}{!}{
    %\input{cooling_different_heat_transfer}}
    \caption{Variation of the heat transfer coefficient, with a current density of $J = 3.5\cdot10\,$A/mm$^2$.}
    \label{fig:investigation_h}
    \end{subfigure}
    \caption{Temperature at position 3 along the winding direction compared for various parameter.}
    \label{fig:investigation}
\end{figure}

%
In machines with a higher power density, the effects of spray cooling will have a greater impact. To illustrate this, the existing machine is simulated with a $\sqrt{10}$ -times larger powering current, which roughly corresponds to a 10 times larger power density and a tenfold larger heat source density, reaching 18.4\,MW/m$^3$. The resulting temperature distribution along the winding direction is shown in Fig.~\ref{fig:temp_axis_5} for selected points. The highest temperature is 410\,K in the center cross-section and 335\,K in the overhang cross-section. Therefore, the highest temperature is still lower than the permitted average winding temperature of the "thermal class 180" of insulating materials \cite{Pyrhonen_2013aa}. The lowest temperature is found at the surface of the overhang, which is only 0.5\,K higher than the spray temperature. \\
To further investigate the cooling effect, the resulting temperature at position 3, which is the hot spot of the winding, is evaluated for different current densities and heat transfer coefficients (see. Fig.~\ref{fig:investigation}). In Fig.~\ref{fig:investigation_J}, a current density between 10 and 40\,A/mm$^2$ is considered. While a current density of 35\,A/mm$^2$ stays within the "thermal class 180" limits, the maximum temperature of 462.9\,K for 40\,A/mm$^2$ exceeds the limits for the insulation class. In Fig.~\ref{fig:investigation_h}, the heat transfer coefficient is evaluated for sprays between 5 and 22.5\,kW/Km$^2$ as well as for air-cooling, which reaches 0.25\,kW/Km$^2$. Additionally, a current density of J=35\,A/mm$^2$ is considered, which was shown above to be the limit for the chosen machine winding. With a temperature difference from the center to the overhang of 88.6\,K for air-cooling, and 84.6\,K and 83.1\,K for spray cooling with a heat transfer coefficient of 5 and 22.5\,kW/Km$^2$, respectively, the cooling behavior varies only by 5\,K. However, the maximum temperature difference between the highest temperature of air and spray-cooling is around 200\,K, which indicates a significant cooling effect using spray cooling. Note that the spray with a heat transfer coefficient of 5\,kW/Km$^2$ is still within the "thermal class 180" limits, which indicates that other spray materials, can be applicable for spray cooling as well.
%41680560
\section{Conclusion}

In this paper, an overview of spray cooling is provided first. The necessary parameters for its description as an impedance boundary with a thermal FE model have been derived. The temperature distribution was simulated with a quasi-3D FE model, which essentially extrudes the 2D cross-sectional FE model along the winding direction to account for the heat flux from the machine's interior to the winding overhang where spray cooling is applied.
It was shown that when applying spray cooling to an initially air-rated winding model, the machine can withstand a tenfold increase in power density. Additionally, the influence of the heat transfer coefficient on the temperature distribution has been evaluated, indicating that other spray fluids may be efficient as well. Therefore, spray cooling is a promising technology to increase the power density of electric machines.

\section*{Acknowledgment}
This work is supported by the joint DFG/FWF Collaborative Research Centre CREATOR (DFG: Project-ID 492661287/TRR 361; FWF: 10.55776/F90, subprojects A03, A04 and B03) at TU Darmstadt, TU Graz and JKU Linz, the Graduate School Computational Engineering at TU Darmstadt, the Athene Young Investigator Program of TU Darmstadt. The authors would like to thank Anouar Belahcen for his valuable advice, and Laura D'Angelo, Greta Ruppert and Max Schaufelberger for their expertise in quasi-3D.

\bibliographystyle{ieeetr}
\bibliography{Bibfile.bib} 

\begin{thebibliography}{10}

\bibitem{rehman2018three}
Z.~Rehman and K.~Seong, ``Three-{D} numerical thermal analysis of electric
  motor with cooling jacket,'' {\em Energies}, vol.~11, no.~1, p.~92, 2018.

\bibitem{fan2010thermal}
J.~Fan, C.~Zhang, Z.~Wang, Y.~Dong, C.~Nino, A.~Tariq, and E.~Strangas,
  ``Thermal analysis of permanent magnet motor for the electric vehicle
  application considering driving duty cycle,'' {\em IEEE Transactions on
  Magnetics}, vol.~46, no.~6, pp.~2493--2496, 2010.

\bibitem{chong2021review}
Y.~C. Chong, D.~Staton, Y.~Gai, H.~Adam, and M.~Popescu, ``Review of advanced
  cooling systems of modern electric machines for e-mobility application,'' in
  {\em 2021 IEEE Workshop on Electrical Machines Design, Control and Diagnosis
  (WEMDCD)}, pp.~149--154, IEEE, 2021.

\bibitem{7194542}
M.~Popescu, D.~Staton, A.~Boglietti, A.~Cavagnino, D.~Hawkins, and J.~Goss,
  ``Modern heat extraction systems for electrical machines - a review,'' in
  {\em 2015 IEEE Workshop on Electrical Machines Design, Control and Diagnosis
  (WEMDCD)}, pp.~289--296, 2015.

\bibitem{en16197006}
D.~Konovalov, I.~Tolstorebrov, T.~M. Eikevik, H.~Kobalava, M.~Radchenko,
  A.~Hafner, and A.~Radchenko, ``Recent developments in cooling systems and
  cooling management for electric motors,'' {\em Energies}, vol.~16, no.~19,
  2023.

\bibitem{9819945}
P.~Shams~Ghahfarokhi, A.~Podgornovs, A.~Kallaste, A.~J. Marques~Cardoso,
  A.~Belahcen, and T.~Vaimann, ``The oil spray cooling system of automotive
  traction motors: The state of the art,'' {\em IEEE Transactions on
  Transportation Electrification}, vol.~9, no.~1, pp.~428--451, 2023.

\bibitem{WANG2022102082}
X.~Wang, B.~Li, K.~Huang, Y.~Yan, I.~Stone, and S.~Worrall, ``Experimental
  investigation on end winding thermal management with oil spray in electric
  vehicles,'' {\em Case Studies in Thermal Engineering}, vol.~35, p.~102082,
  2022.

\bibitem{FARSANE20001321}
K.~Farsane, P.~Desevaux, and P.~K. Panday, ``Experimental study of the cooling
  of a closed type electric motor,'' {\em Applied Thermal Engineering},
  vol.~20, no.~14, pp.~1321--1334, 2000.

\bibitem{ghahfarokhi2020development}
P.~S. Ghahfarokhi, A.~Kallaste, A.~Podgornovs, A.~Belahcen, and T.~Vaimann,
  ``Development of analytical thermal analysis tool for synchronous reluctance
  motors,'' {\em IET Electric Power Applications}, vol.~14, no.~10,
  pp.~1828--1836, 2020.

\bibitem{shams2020analytical}
P.~Shams~Ghahfarokhi, A.~Kallaste, A.~Belahcen, and T.~Vaimann, ``Analytical
  thermal model and flow network analysis suitable for open self-ventilated
  machines,'' {\em IET Electric Power Applications}, vol.~14, no.~6,
  pp.~929--936, 2020.

\bibitem{chen2022application}
H.~Chen, X.-h. Ruan, Y.-h. Peng, Y.-l. Wang, and C.-k. Yu, ``Application status
  and prospect of spray cooling in electronics and energy conversion
  industries,'' {\em Sustainable energy technologies and assessments}, vol.~52,
  p.~102181, 2022.

\bibitem{8848870}
C.~Liu, Z.~Xu, D.~Gerada, J.~Li, C.~Gerada, Y.~C. Chong, M.~Popescu, J.~Goss,
  D.~Staton, and H.~Zhang, ``Experimental investigation on oil spray cooling
  with hairpin windings,'' {\em IEEE Transactions on Industrial Electronics},
  vol.~67, no.~9, pp.~7343--7353, 2020.

\bibitem{10239016}
F.~Hoffmann, J.~Bender, M.~Parche, T.~Wetzel, and M.~Doppelbauer, ``Local heat
  transfer coefficient measurements on shaft spray cooled end windings,'' in
  {\em 2023 IEEE International Electric Machines \& Drives Conference (IEMDC)},
  pp.~1--7, 2023.

\bibitem{zhang2021thermal}
F.~Zhang, D.~Gerada, Z.~Xu, C.~Liu, H.~Zhang, T.~Zou, Y.~C. Chong, and
  C.~Gerada, ``A thermal modeling approach and experimental validation for an
  oil spray-cooled hairpin winding machine,'' {\em IEEE Transactions on
  Transportation Electrification}, vol.~7, no.~4, pp.~2914--2926, 2021.

\bibitem{zhao2023parameter}
A.~Zhao, C.~Duwig, C.~Liu, D.~Gerada, and M.~Leksell, ``Parameter study for oil
  spray cooling on endwindings of electric machines via eulerian--lagrangian
  simulation,'' {\em Applied Thermal Engineering}, vol.~235, p.~121281, 2023.

\bibitem{el2014advanced}
A.~M. El-Refaie, J.~P. Alexander, S.~Galioto, P.~B. Reddy, K.-K. Huh,
  P.~de~Bock, and X.~Shen, ``Advanced high-power-density interior permanent
  magnet motor for traction applications,'' {\em IEEE Transactions on Industry
  Applications}, vol.~50, no.~5, pp.~3235--3248, 2014.

\bibitem{dong2021performance}
H.~Dong, L.~Ruan, Y.~Wang, J.~Yang, F.~Liu, and S.~Guo, ``Performance of
  air/spray cooling system for large-capacity and high-power-density motors,''
  {\em Applied Thermal Engineering}, vol.~192, p.~116925, 2021.

\bibitem{5382944}
Z.~Li, J.~Guo, D.~Fu, G.~Gu, and B.~Xiong, ``Research on heat transfer of
  spraying evaporative cooling technique for large electrical machine,'' in
  {\em 2009 International Conference on Electrical Machines and Systems},
  pp.~1--4, 2009.

\bibitem{10.1260/1757-482x.5.4.239}
M.~R. Guechi, P.~Désévaux, and P.~Baucour, ``On the numerical and
  experimental study of spray cooling,'' {\em The Journal of Computational
  Multiphase Flows}, vol.~5, pp.~239--249, 2013.

\bibitem{doi:10.1177/1757482X16653895}
M.~Guechi, P.~Desevaux, P.~Baucour, C.~Espanet, R.~Brunel, and M.~Poirot,
  ``Spray cooling of electric motor coil windings,'' {\em The Journal of
  Computational Multiphase Flows}, vol.~8, no.~2, pp.~95--100, 2016.

\bibitem{albrecht2013laser}
H.-E. Albrecht, N.~Damaschke, M.~Borys, and C.~Tropea, {\em Laser Doppler and
  phase Doppler measurement techniques}.
\newblock Springer Science \& Business Media, 2013.

\bibitem{tropea2011optical}
C.~Tropea, ``Optical particle characterization in flows,'' {\em Annual Review
  of Fluid Mechanics}, vol.~43, no.~1, pp.~399--426, 2011.

\bibitem{mudawar1994universal}
I.~Mudawar and T.~A. Deiters, ``A universal approach to predicting temperature
  response of metallic parts to spray quenching,'' {\em International journal
  of heat and mass transfer}, vol.~37, no.~3, pp.~347--362, 1994.

\bibitem{hall1995experimental}
D.~D. Hall and I.~Mudawar, ``Experimental and numerical study of quenching
  complex-shaped metallic alloys with multiple, overlapping sprays,'' {\em
  International journal of heat and mass transfer}, vol.~38, no.~7,
  pp.~1201--1216, 1995.

\bibitem{tilton1994liquid}
D.~E. Tilton, D.~A. Kearns, and C.~L. Tilton, ``Liquid nitrogen spray cooling
  of a simulated electronic chip,'' in {\em Advances in cryogenic engineering},
  pp.~1779--1786, Springer, 1994.

\bibitem{yin2022spray}
J.~Yin, S.~Wang, X.~Sang, Z.~Zhou, B.~Chen, P.~Thrassos, A.~Romeos, and
  A.~Giannadakis, ``Spray cooling as a high-efficient thermal management
  solution: a review,'' {\em Energies}, vol.~15, no.~22, p.~8547, 2022.

\bibitem{huddle2000thermal}
J.~J. Huddle, L.~C. Chow, S.~Lei, A.~Marcos, D.~P. Rini, S.~J. Lindauer,
  M.~Bass, and P.~J. Delfyett, ``Thermal management of diode laser arrays,'' in
  {\em Sixteenth Annual IEEE Semiconductor Thermal Measurement and Management
  Symposium (Cat. No. 00CH37068)}, pp.~154--160, IEEE, 2000.

\bibitem{liu2019applications}
R.~Liu, L.~Zhang, and X.~Zhang, ``Applications of spray cooling technology in
  aerospace field,'' in {\em IOP Conference Series: Materials Science and
  Engineering}, vol.~470, p.~012020, IOP Publishing, 2019.

\bibitem{torres1999estimation}
J.~H. Torres, J.~S. Nelson, B.~S. Tanenbaum, T.~E. Milner, D.~M. Goodman, and
  B.~Anvari, ``Estimation of internal skin temperatures in response to cryogen
  spray cooling: implications for laser therapy of port wine stains,'' {\em
  IEEE Journal of selected topics in Quantum Electronics}, vol.~5, no.~4,
  pp.~1058--1066, 1999.

\bibitem{zhang2022advanced}
T.~Zhang, Z.~Mo, X.~Xu, X.~Liu, H.~Chen, Z.~Han, Y.~Yan, and Y.~Jin, ``Advanced
  study of spray cooling: from theories to applications,'' {\em Energies},
  vol.~15, no.~23, p.~9219, 2022.

\bibitem{kim2007spray}
J.~Kim, ``Spray cooling heat transfer: The state of the art,'' {\em
  International Journal of Heat and Fluid Flow}, vol.~28, no.~4, pp.~753--767,
  2007.

\bibitem{liang2017review}
G.~Liang and I.~Mudawar, ``Review of spray cooling--part 1: Single-phase and
  nucleate boiling regimes, and critical heat flux,'' {\em International
  Journal of Heat and Mass Transfer}, vol.~115, pp.~1174--1205, 2017.

\bibitem{Breitenbach2018aa}
J.~Breitenbach, I.~V. Roisman, and C.~Tropea, ``From drop impact physics to
  spray cooling models: a critical review,'' {\em Experiments in Fluids},
  vol.~59, pp.~1--21, 2018.

\bibitem{moreira2010advances}
A.~L.~N. Moreira, A.~S. Moita, and M.~R. Panao, ``Advances and challenges in
  explaining fuel spray impingement: How much of single droplet impact research
  is useful?,'' {\em Progress in energy and combustion science}, vol.~36,
  no.~5, pp.~554--580, 2010.

\bibitem{yarin2017collision}
A.~L. Yarin, I.~V. Roisman, and C.~Tropea, {\em Collision phenomena in liquids
  and solids}.
\newblock Cambridge University Press, 2017.

\bibitem{cossali1997impact}
G.~E. Cossali, A.~Coghe, and M.~Marengo, ``The impact of a single drop on a
  wetted solid surface,'' {\em Experiments in fluids}, vol.~22, no.~6,
  pp.~463--472, 1997.

\bibitem{mundo1995droplet}
C.~Mundo, M.~Sommerfeld, and C.~Tropea, ``Droplet-wall collisions: experimental
  studies of the deformation and breakup process,'' {\em International journal
  of multiphase flow}, vol.~21, no.~2, pp.~151--173, 1995.

\bibitem{van2010dynamics}
N.~P. {v}an Hinsberg, M.~Budakli, S.~G{\"o}hler, E.~Berberovi{\'c}, I.~V.
  Roisman, T.~Gambaryan-Roisman, C.~Tropea, and P.~Stephan, ``Dynamics of the
  cavity and the surface film for impingements of single drops on liquid films
  of various thicknesses,'' {\em Journal of colloid and interface science},
  vol.~350, no.~1, pp.~336--343, 2010.

\bibitem{stumpf2022drop}
B.~Stumpf, J.~Hussong, and I.~V. Roisman, ``Drop impact onto a substrate wetted
  by another liquid: Flow in the wall film,'' {\em Colloids and Interfaces},
  vol.~6, no.~4, p.~58, 2022.

\bibitem{DAngelo_2020}
L.~A.~M. D’Angelo, J.~Christ, and H.~De~Gersem, ``Quasi-3d discretization of
  thermal hot-spot propagation in superconducting models,'' {\em IEEE
  Transactions on Applied Superconductivity}, vol.~30, no.~4, pp.~1--5, 2020.

\bibitem{WEN2019431}
T.~Wen, Y.~Luo, W.~He, W.~Gang, and L.~Sheng, ``Development of a novel quasi-3d
  model to investigate the performance of a falling film dehumidifier with cfd
  technology,'' {\em International Journal of Heat and Mass Transfer},
  vol.~132, pp.~431--442, 2019.

\bibitem{https://doi.org/10.1049/iet-epa.2015.0491}
C.~Liu, J.~Ruan, W.~Wen, R.~Gong, and C.~Liao, ``Temperature rise of a dry-type
  transformer with quasi-3d coupled-field method,'' {\em IET Electric Power
  Applications}, vol.~10, no.~7, pp.~598--603, 2016.

\bibitem{Driendl_2022aa}
N.~Driendl, F.~Pauli, and K.~Hameyer, ``Influence of ambient conditions on the
  qualification tests of the interturn insulation in low-voltage electrical
  machines,'' {\em IEEE Transactions on Industrial Electronics}, vol.~69,
  no.~8, pp.~7807--7816, 2022.

\bibitem{Eickhoff_2020aa}
H.~T. Eickhoff, {\em Fault-Tolerant Operation of Three-Phase Squirrel Cage
  Induction Machine Drives after Inverter Open Circuit Faults}.
\newblock Dissertation, Graz University of Technology, 2020.

\bibitem{bergfried2023aa}
C.~Bergfried, Y.~Sp{\"a}ck-Leigsnering, R.~Seebacher, H.~Eickhoff, and
  A.~Muetze, ``Thermal finite element modeling and simulation of a
  squirrel-cage induction machine,'' {\em International Journal of Applied
  Electromagnetics and Mechanics}, no.~Preprint, pp.~1--14, 2023.

\bibitem{10700088}
L.~Blumrich, C.~Bergfried, A.~Galetzka, H.~De~Gersem, R.~Seebacher, A.~Mütze,
  and Y.~Späck-Leigsnering, ``Thermal model calibration of a squirrel-cage
  induction machine,'' in {\em 2024 International Conference on Electrical
  Machines (ICEM)}, pp.~1--7, 2024.

\bibitem{Chemieingenieurwesen_2010aa}
``{VDI} heat atlas,'' Springer, 2010.

\bibitem{Bundschuh_2023ab}
J.~Bundschuh, M.~G. Ruppert, and Y.~Sp{\"a}ck-Leigsnering, ``Pyrit: A finite
  element based field simulation software written in python,'' {\em COMPEL},
  2023.

\bibitem{Comsol}
``Comsol multiphysics,'' (Stockholm, Sweden), {COMSOL AB}.

\bibitem{meschede2015gerthsen}
D.~Meschede and C.~Gerthsen, {\em Gerthsen Physik}.
\newblock Springer-Lehrbuch, Springer, 2015.

\bibitem{Fletcher.1984}
C.~A.~J. Fletcher, {\em Galerkin Finite-Element Methods}, pp.~86--154.
\newblock Berlin, Heidelberg: Springer Berlin Heidelberg, 1984.

\bibitem{9233257}
D.~Paloschi, K.~A. Bronnikov, S.~Korganbayev, A.~A. Wolf, A.~Dostovalov, and
  P.~Saccomandi, ``3d shape sensing with multicore optical fibers:
  Transformation matrices versus frenet-serret equations for real-time
  application,'' {\em IEEE Sensors Journal}, vol.~21, no.~4, pp.~4599--4609,
  2021.

\bibitem{kyriopoulos2010gravity}
O.~N. Kyriopoulos, {\em Gravity effect on liquid film hydrodynamics and spray
  cooling}.
\newblock PhD thesis, Technische Universit{\"a}t Darmstadt, Darmsdadt, Germany,
  2010.

\bibitem{gambaryan2007gravity}
T.~Gambaryan-Roisman, O.~Kyriopoulos, I.~Roisman, P.~Stephan, and C.~Tropea,
  ``Gravity effect on spray impact and spray cooling,'' {\em Microgravity
  Science and Technology}, vol.~19, pp.~151--154, 2007.

\bibitem{Pyrhonen_2013aa}
J.~Pyrh\"onen, T.~Jokinen, and V.~Hrabovcov\'a, {\em Design Rotating Electrical
  Machines}.
\newblock John Wiley \& Sons, 2013.

\end{thebibliography}

\begin{acronym}
        \acro{1d}[1D]{one-dimensional}
	\acro{2d}[2D]{two-dimensional}
        \acro{3d}[3D]{three-dimensional}
        \acro{bc}[BC]{boundary condition}
        \acro{fe}[FE]{finite-element}
        \acro{cfd}[CFD]{computational fluid dynamics}
\end{acronym}
\end{document}